\documentstyle[12pt]{article}

\begin{document}
\vspace{-2.0cm}
\bigskip
\begin{center}
{\Large \bf
Polarization Vectors,  Doublet Structure and Wigner's Little Group 
in  Planar Field Theory}
\end{center}
\vskip .8 true cm
\begin{center}
{\bf Rabin Banerjee}\footnote{rabin@boson.bose.res.in},
{\bf Biswajit Chakraborty}\footnote{biswajit@boson.bose.res.in} and
{\bf Tomy Scaria}\footnote{tomy@boson.bose.res.in}

\vskip 1.0 true cm

S. N. Bose National Centre for Basic Sciences \\
JD Block, Sector III, Salt Lake City, Calcutta -700 098, India.
        
\end{center}
\bigskip

\centerline{\large \bf Abstract}
We  establish the equivalence of the Maxwell-Chern-Simons-Proca
 model  to a doublet of Maxwell-Chern-Simons models at the level of 
polarization vectors 
of the basic fields using both Lagrangian and Hamiltonian formalisms. The 
analysis reveals a $U(1)$ invariance of the polarization vectors in the 
momentum space. Its implications are discussed.
We also study the role of Wigner's little group as a generator of gauge 
transformations in three space-time dimensions. \\
PACS No.(s) 11.10.Kk, 11.15.-q
\newpage
\section{Introduction}
Recently there has been a great deal of interest in the study of 2+1 dimensional field theoretical models \cite{seigel,djt}. Among the various applications mention
 may be made of its utility for studying high temperature asymptotics of
4-d models and different effects involving planar systems in condensed matter 
\cite{kallosh,wilczek}. 

Some distinctive properties of these theories are linked to the co-existence of gauge invariance with mass, which is not due to any loop effects. Thus in contrast to the Schwinger model, gauge invariant mass occurs already at the classical
level. The mass is introduced by adding a Chern - Simons term to the bare 
(Maxwell) action, leading to a topologically massive gauge theory(Maxwell-Chern-Simons(MCS) theory)\cite{seigel,djt}.   

It might appear that lower dimensional theories would be simpler than their
realistic four dimensional versions. However, it must be remembered that 
low dimensional field theories suffer from severe infra red singularities. 
A possible approach to tackle this situation is to introduce a regulator in 
the form of a usual mass term to the original topologically massive gauge 
theory(Maxwell-Chern-Simons-Proca(MCSP) theory) and study the vanishing mass
limit\cite{tyutin}.

Now, it is intriguing to note that the topologically massive theory augmented
by the usual mass term can be regarded as the embedding of a doublet of 
topologically massive gauge theories\cite{bk,bkm}. In view of the comments in the
preceding para, we feel that this connection deserves a more detailed study.
Earlier, this was done at the level of the basic fields in the two theories.
In the present paper, we shall pursue this mapping at the more fundamental
level involving polarization vectors of these fields. This is all the more 
important since proper evaluation of these vectors is crucial for reduction
formulae and the study of the massless limit of the MCSP theory \cite{tyutin}.

In section 2, we calculate the polarization vectors in the Lagrangian formalism
using  different approaches. There is a doublet structure in the MCSP theory
which is identical with a pair of MCS theories having opposite helicities.
An Abelian ($U(1)$) invariance is found in the polarization vectors
 in the momentum
space. This is exploited to establish the equivalence between apparently
different expressions for these vectors. In section 3, the calculation 
of the polarization vectors is done in the Hamiltonian approach using Dirac's
constrained formalism. The equivalence with the results in the Lagrangian
approach is shown. Section 4 is an aside on Wigner's little group in
2+1 dimensions. The explicit group structure is obtained. The role of the 
little group as a generator of gauge transformation is elaborated. Finally,
we show that the spin of the MCS quanta may be speculated from these 
properties. Section 5 contains the concluding remarks.

\section{Polarization vectors in the  
Lagrangian formalism}
Here a detailed calculation of the polarization vectors of Maxwell - Chern - 
Simons(MCS) theory will be presented. Apart from reviewing the standard analysis
\cite{greiner,girotti} in the Lorentz gauge, a completely gauge independent 
analysis will also be discussed. The expressions found by this approach
match with the Lorentz gauge expressions.
\subsection{Maxwell - Chern - Simons theory}
\begin{sloppypar}
We first review the calculation of the polarization vectors in the Maxwell-Chern-Simons theory pointing out the differences from  the corresponding 
analysis for Maxwell theory. \\
The MCS Lagrangian in 2+1 dimensions is given by
\end{sloppypar}
\bigskip
$$ {\cal L} = -{\frac{1}{4}}{F^{\mu \nu}F_{\mu \nu}} + {\frac{\vartheta}{2}}{\epsilon}^{\mu \nu \lambda}A_{\mu}{\partial}_{\nu}A_{\lambda}. \eqno{(2.1.1)} $$ 
This is the well known topological gauge theory with a single mode of mass
$|\vartheta|$ and spin $\frac{\vartheta}{|\vartheta|}$ \cite{djt}.
The corresponding equation of motion is given by,
\bigskip
$$ -{\partial}^{\nu}F_{\mu \nu} + {\vartheta}\epsilon _{\mu \nu \lambda }{\partial}^{\nu}A^{\lambda} = 0. \eqno {(2.1.2)} $$
Imposing the Lorentz gauge,
\bigskip
$$ {\partial}_{\mu}A^{\mu} = 0 \eqno {(2.1.3)}  $$
the above equation reduces to,

$$ \left( {\Box}g^{\mu \nu} + {\vartheta \epsilon ^{\mu \lambda \nu }}{\partial}_{\lambda} \right) A_{\nu} = 0. \eqno {(2.1.4)} $$
Substituting the solution for the negative energy component in terms of the polarization vector,
\bigskip
$$ A^{\mu} = {\eta}^{\mu}(k)\exp(ik.x) \eqno {(2.1.5)} $$
in the above two equations, gives, respectively,
\bigskip
$$ k_{\mu}{\eta}^{\mu} = 0 \eqno {(2.1.6)} $$
and 
$$ {\Sigma}_{(MCS)}^{\mu \nu}{\eta}_{\nu}(k) = 0. \eqno {(2.1.7)} $$
where,
\bigskip
$$ {\Sigma}_{(MCS)}^{\mu \nu} = -k^2g^{\mu \nu} + i{\vartheta \epsilon ^{\mu \lambda \nu}}k_{\lambda} \eqno {(2.1.8).} $$
For a non trivial solution,  
\bigskip
$$ \det {\Sigma}_{(MCS)} = -k^4(k^2 - {\vartheta}^2) = 0 \eqno {(2.1.9)} $$
so that, 

$$  k^2 = 0 \eqno {(2.1.10a)} $$
   or 
$$  k^2 = {\vartheta}^2. \eqno {(2.1.10b)}   $$ 
When $k^2 = 0$, the solution is,
\bigskip
$$ {\eta}^{\mu}(k) = k^{\mu}f(k) $$
where $f(k)$ is an arbitrary function. 
Therefore massless excitations are pure gauge artefacts, which may be ignored. 

Now consider the case $k^2 = {\vartheta}^2$, which implies quanta of mass 
$|\vartheta|$. This enables a passage to the rest frame $k^{\mu} = (|\vartheta|,0,0)$. Then 
the equation of motion (2.1.7) yields,
\bigskip 
$$  -{\vartheta}^2{\eta}_0({\bf 0}) = 0 $$
$$  {\vartheta}^2{\eta}_1({\bf 0}) + i{\vartheta}(-|{\vartheta}|){\eta}_2({\bf 0}) = 0 $$
$$ {\vartheta}^2{\eta}_2({\bf 0}) + i{\vartheta}|{\vartheta}|{\eta}_1({\bf 0}) = 0 $$
where $ {\eta}({\bf 0}) $ stands for MCS polarization in the rest frame.
The above set yields,
\bigskip
$$ {\eta}^0({\bf 0}) = 0 \eqno {(2.1.11a)} $$
$$   {\eta}^2({\bf 0}) = -i{\frac{\vartheta}{|\vartheta|}}{\eta}^1({\bf 0}). \eqno {(2.1.11b)} $$
Therefore in the rest frame the polarization vector is given by, 
\bigskip
$$ {\eta}^{\mu}({\bf 0}) = \left( 0, {\eta}^1({\bf 0}), -i{\frac{\vartheta}{|\vartheta|}}{\eta}^1({\bf 0}) \right).  \eqno {(2.1.12)} $$
The above expressions are determined modulo a normalization factor. This can be
fixed from the condition, 
\bigskip
$$ \eta^{\ast \mu} ({\bf 0}) \eta_{\mu}({\bf 0}) = -1. \eqno{(2.1.13)} $$ 
following essentially from the space-like nature of the vector $\eta^\mu$. An
important point of distinction from the Maxwell case is that $\eta^\mu$ has
complex entries so that the norm is real. Furthermore, the normalization 
condition reveals a $U(1)$  invariance in the expression for $\eta^\mu$;
i.e., if $\eta^\mu$ is a solution, then $e^{i\phi} \eta^\mu$ is also a solution.
 This observation will be used later on to show the equivalence among different
forms for $\eta^\mu$.

The normalization condition fixes $|{\eta}^1({\bf 0})|^2 = \frac{1}{2}$. Hence,
\bigskip 
$$ {\eta}^{\mu}({\bf 0}) = \frac{1}{\sqrt{2}}\left(0, 1, -i{\frac{\vartheta}{|\vartheta|}}\right)  \eqno {(2.1.14)}  $$

Now we present another, gauge independent derivation of this result where only the symmetries of the
model are used.  Consider again the equation (2.1.2) and assume solutions of the form (2.1.5).
Substituting (2.1.5) in (2.1.2) yields,
\bigskip
$$ \eta_{\nu} k^{\nu}k^{\mu} - k^2 \eta^{\mu} + i\vartheta \epsilon^{\mu \nu \lambda} \eta_{\lambda} k_{\nu}
= 0  \eqno {(2.1.15)} $$
The two  possibilities for $k^2$, corresponding to massless or massive modes  are, 
$$ (i) \hskip 0.5cm  k^2 = 0  \eqno {(2.1.16a)} $$
$$ (ii) \hskip 0.5cm k^2 \not= 0 \eqno {(2.1.16b)} $$
We first take up the case (i) i.e. $k^2 = 0$.
Using (2.1.16a) in (2.1.15) gives,
$$ k^{\mu}(\eta \cdot k) = -i \vartheta \epsilon^{\mu \nu \lambda} k_{\nu} \eta_{\lambda} $$ 
Multiplying both sides with $\epsilon_{\mu \alpha \beta} k^{\alpha}$ one arrives at,
 $$ 0 = i\vartheta k_{\beta}(\eta \cdot k) $$
which implies, 
$$ \eta \cdot k = 0. \eqno {(2.1.17)} $$
Using (2.1.16a)  and (2.1.17) in (2.1.15), we get $\eta^{\mu} = f(k)k^{\mu}$ which, as mentioned
earlier, shows that massless excitations are pure gauge artefacts.

Next we consider (2.1.16b). Since $k^2 \not= 0$, from (2.1.15) we have,
\bigskip
$$ \eta^{\mu} = \frac{1}{k^2}\left[ (\eta \cdot k)k_{\mu} + i\vartheta \epsilon_{\mu \nu \lambda} k^{\nu} \eta^{\lambda} \right].  \eqno {(2.1.18)} $$ 
 and we are allowed to go to a rest frame where $k^{\mu} = (m, 0, 0)$ and $k^2 = m^2$.
Let $\eta^{\mu}$ in this frame be given by, 
\bigskip
$$ \eta^{\mu}({\bf 0}) = \left( \eta^0 ({\bf 0}), \eta^1 ({\bf 0}), \eta^2 {\bf 0})\right). $$
Then (2.1.18) gives, 
\bigskip
$$ \eta^1({\bf 0}) = \frac{i\vartheta}{m} \eta^2 ({\bf 0}) \eqno {(2.1.19a)} $$
and 
$$ \eta^2({\bf 0}) = - \frac{i\vartheta}{m} \eta^1 ({\bf 0}). \eqno {(2.1.19b)} $$
Substituting for $\eta^2({\bf 0})$ from (2.1.19b) in (2.1.19a) gives,
\bigskip
$$ \vartheta^2 = m^2 $$
from which it follows,  
\bigskip
$$ m = | \vartheta |. \eqno {(2.1.20)} $$
From the gauge invariance of the model it  follows that 
${\eta}^0({\bf 0})$ can be set equal to zero.
Therefore from (2.1.19a, b) and  (2.1.20), we we reproduce the earlier result(2.1.11).
It is important to note that the result is compatible with the covariance condition (2.1.6)
 although it was not used explicitly in the analysis. 
This is an important difference from the treatment of the Maxwell theory where
the polarization vector does not satisfy this condition. It has, therefore, to
be enforced by hand. This is the Gupta - Bleuler technique where the physical
sector is projected out by using such a condition. The origin of this difference is the curious property of the MCS theory where gauge invariance coexists with
massive modes.

It is now straightforward to calculate the polarization vector in a moving
 frame by giving a Lorentz boost to the result in the rest frame,
\bigskip
$$ \left( \begin{array}{c}
  {\eta}^0(k) \\ {\eta}^1( k) \\ {\eta}^2(k) \end{array} \right) = \left( \begin{array}{ccc}
   {\gamma} & {\gamma}{\beta}^1 & {\gamma}{\beta}^2	 \\
	{\gamma}{\beta}^1 & 1 + \frac{({\gamma} -1)({\beta}^1)^2}{(\vec{\beta})^2} & \frac{({\gamma} -1){\beta}^1{\beta}^2}{(\vec{\beta})^2} \\
	{\gamma}{\beta}^2 & \frac{({\gamma} -1){\beta}^1{\beta}^2}{(\vec{\beta})^2} & 1 + \frac{({\gamma} -1)({\beta}^2)^2}{(\vec{\beta})^2}
    \end{array} \right)\left( \begin{array}{c}
  {\eta}^0({\bf 0}) \\ {\eta}^1({\bf 0}) \\ {\eta}^2({\bf 0}) \end{array} \right)  \eqno {(2.1.21)}  $$ 
where $\vec{\beta} = \frac{\bf k}{k^0}$ and $\gamma = \frac{k^0}{|\vartheta|}$.
The ensuing polarization vector is given by,
\bigskip
$$ {\eta}^{\mu}(k) = \left( \frac{{\vec{\eta}({\bf 0})}\cdot{\bf k}}{|\vartheta|},	\vec{\eta}({\bf 0}) + \frac{{\vec{\eta}({\bf 0})}\cdot{\bf k}}{(k^0 + |\vartheta|)|\vartheta|} {\bf k} \right) \eqno {(2.1.22a)} $$
where $\vec{\eta}({\bf 0})$ stands for the space part of the vector in (2.1.14). Thus,
\bigskip
$$ {\eta}^{\mu}(k) = 
\left( \frac{k^1 - i{\frac{\vartheta}{|\vartheta|}}k^2}{{\sqrt{2}}{|\vartheta|}},  
\frac{1}{\sqrt{2}} + \frac{k^1 - i{\frac{\vartheta}{|\vartheta|}}k^2}{{\sqrt{2}}{(k^0 + |\vartheta|)|\vartheta|}}k^1, -i\frac{\vartheta}{\sqrt{2}|\vartheta|} + 
\frac{k^1 - i{\frac{\vartheta}{|\vartheta|}
}k^2}{{\sqrt{2}}{(k^0 + |\vartheta|)|\vartheta|}}k^2 \right) \eqno {(2.1.22b)} $$
which agrees with the expression given in \cite{girotti}.
\subsection{Maxwell - Chern - Simons - Proca theory}
The Maxwell-Chern-Simons-Proca(MCSP) Lagrangian is given by,
\bigskip
$$ {\cal L} = -{\frac{1}{4}}{F^{\mu \nu}F_{\mu \nu}} + {\frac{\theta}{2}}{\epsilon}^{\mu \nu \lambda}A_{\mu}{\partial}_{\nu}A_{\lambda} + {\frac{m^2}{2}}A^{\mu}A_{\mu}. \eqno {(2.2.1)} $$
The equation of motion is, 
$$ -{\partial}^{\nu}F_{\mu \nu} + {\theta}\epsilon _{\mu \nu \lambda}{\partial}^{\nu}A^{\lambda} +m^2A_{\mu} = 0 \eqno {(2.2.2)}$$
which satisfies the  transversality condition
 ${\partial}_{\mu}A^{\mu} = 0.$
Using this, the equation of motion simplifies to,
\bigskip
$$ \left[({\Box} + m^2)g^{\mu \nu} + {\theta}{\epsilon}^{\mu \lambda \nu}{\partial}_{\lambda} \right]A_{\nu} = 0 \eqno {(2.2.3)} $$
Substitution of the solution $A^{\mu} = {\varepsilon}^{\mu}(k){\exp}(ik.x)$ yields,
\bigskip
$$ \left[(-k^2 + m^2)g^{\mu \nu} + i{\theta \epsilon ^{\mu \lambda \nu}}k_{\lambda}\right]{\varepsilon}_{\nu} = 0. \eqno {(2.2.4)} $$
From the transversality relation we get $k_{\mu}{\varepsilon}^{\mu} = 0.$
Let us define,
\bigskip
$$ {\Sigma}^{\mu \nu}_{(MCSP)} = (-k^2 + m^2)g^{\mu \nu} + i{\theta \epsilon ^{\mu \lambda \nu}}k_{\lambda} \eqno {(2.2.5)} $$
Then the equation of motion can be written as,   
\bigskip
$$ {\Sigma}^{\mu \nu}_{(MCSP)}{\varepsilon}_{\nu} = 0. \eqno {(2.2.6)} $$
For ${\varepsilon}_{\nu}$ to have a non trivial solution the determinant of 
${\Sigma}_{(MCSP)}$ should vanish. That is,
\bigskip
$$ (-k^2 + m^2)\left[(-k^2 + m^2)^2 - {\theta}^2k^2\right] = 0 \eqno {(2.2.7)} $$
This implies, either, 
\bigskip
$$ -k^2 + m^2 = 0 \eqno {(2.2.8a)} $$
or
\bigskip
$$ (-k^2 + m^2)^2 - {\theta}^2k^2 = 0. \eqno {(2.2.8b)} $$
Using (2.2.8a) in (2.2.4), it follows that the solution must have the form, ${\varepsilon}^{\mu}(k) = k^{\mu}f(k)$,
which is however incompatible with the transeversality relation and is therefore ignored. The second
case leads to
\bigskip
$$ k^2 = \theta_{\pm}^2 \eqno {(2.2.9)} $$
where,
\bigskip 
$$  \theta_{\pm} = \sqrt{\frac{2m^2 + {\theta}^2 \pm \sqrt{{\theta}^4 + 4m^2{\theta}^2}}{2}} = \sqrt{\frac{\theta^2}{4} + m^2} \pm \frac{\theta}{2}. \eqno {(2.2.10)} $$
Two useful relations follow from this identification,
\bigskip
$$ \theta = {\theta}_+ - {\theta}_-  \eqno {(2.2.11a)} $$
and
$$ m^2 = {\theta}_+{\theta}_-. \eqno {(2.2.11b)} $$
We use the notation ${\varepsilon}_{\pm}(k_{\pm})$ for the polarization vectors corresponding to $k^2 = {\theta_{\pm}}^2$ and let ${\varepsilon}_{\pm}({\bf 0})$ denote the polarization vectors in the rest frame.          
Taking the rest frames to be the ones in which $(k^0,0,0) = (|{\theta}_{\pm}|,0,0)$ we have from the equation of motion (2.2.4),
\bigskip
$$ (m^2 - {{\theta}_{\pm}}^2){\varepsilon}_{\pm 0}({\bf 0}) = 0 $$ 
$$ -(m^2 - {{\theta}_{\pm}}^2){\varepsilon}_{\pm 1}({\bf 0}) - i{\theta}{\theta}_{\pm}{\varepsilon}_{\pm 2}({\bf 0}) = 0 $$
$$ -(m^2 - {{\theta}_{\pm}}^2){\varepsilon}_{\pm 2}({\bf 0}) + i{\theta}{\theta}_{\pm}{\varepsilon}_{\pm 1}({\bf 0}) = 0 $$  
where ${\varepsilon}_{\pm}^{\mu} = ({\varepsilon}_{\pm}^0, {\varepsilon}_{\pm}^1, {\varepsilon}_{\pm}^2).$
From the above set of equations we arrive at, 
\bigskip
$$ {\varepsilon}_{\pm 0}({\bf 0}) = 0  \eqno {(2.2.12a)} $$
$$ {\varepsilon}_{\pm 2}({\bf 0}) = \frac{i{\theta}{\theta}_{\pm}}{m^2 - {{\theta}_{\pm}}^2} {\varepsilon}_{\pm 1}({\bf 0}) = \mp i{\varepsilon}_{\pm 1}({\bf 0}) \eqno {(2.2.12b)} $$
where the connection among various parameters has been used.
Using a  normalization condition analogous to (2.1.13)
\bigskip
$$ {\varepsilon}_{\pm}^{\ast \mu}({\bf 0}){\varepsilon}_{\pm \mu}({\bf 0}) = -1 \eqno {(2.2.13)}$$
gives,
\bigskip 
$$ |{\varepsilon}_{\pm 1}({\bf 0})|^2 =\frac{1}{2}. $$
Hence, 
\bigskip 
$$ {\varepsilon}_{\pm}^{\mu}({\bf 0}) = \frac{1}{\sqrt{2}}\left(0, 1, \mp i\right). \eqno {(2.2.14)} $$
The transversality condition $k_{\mu}{\varepsilon}^{\mu} = 0 $ is preserved
which acts as a consistency check.  
\bigskip
The polarization vectors in a moving frame corresponding to the two massive modes with masses ${\theta}_{\pm}$ are easily found, as before, by giving a
Lorentz boost, 
\bigskip
$$ {\varepsilon}_{\pm}^{\mu}(k_{\pm}) = 
 \left(\frac{k^1 \mp i
k^2}{{\sqrt{2}
}{{\theta}_{\pm}}}, \frac{1}{\sqrt{2}} + \frac{k^1 \mp i
k^2}{{\sqrt{2}}{(k^0_{\pm} + {\theta}_{\pm}){\theta}_{\pm}}}k^1, \mp \frac{i}{\sqrt{2}} +\frac{k^1 \mp ik^2}{{\sqrt{2}}{(k^0_{\pm} + {\theta}_{\pm}){\theta}_{\pm}}}k^2 \right) \eqno {(2.2.15)}  $$
The pair of polarization vectors are related by the parity transformation in
two space dimensions $k^1 \rightarrow k^1, k^2 \rightarrow  -k^2$ augmented by $ k^0 _+ \rightarrow k^0 _-$(which also implies   $\theta_+ \rightarrow \theta_-$),
\bigskip
$$ {\varepsilon}_+^0(k^0_+, k^1, k^2) = {\varepsilon}_-^0(k^0_- \rightarrow k^0_+, k^1 \rightarrow k^1, k^2 \rightarrow  -k^2) $$
$$ {\varepsilon}_+^1(k^0_+, k^1, k^2) = {\varepsilon}_-^1(k^0_- \rightarrow k^0_+, k^1 \rightarrow k^1
, k^2 \rightarrow  -k^2) \eqno {(2.2.16)} $$
$$ {\varepsilon}_+^2(k^0_+, k^1, k^2) =  -{\varepsilon}_-^2(k^0_- \rightarrow k^0_+, k^1 \rightarrow k^1
, k^2 \rightarrow  -k^2). $$
Also, the pair is related by complex conjugation,
\bigskip
$$ {\varepsilon}_+^{\mu}(k_+) = {\varepsilon}_-^{\ast \mu}(k_-). \eqno {(2.2.17)} $$

Now it may be pointed out that the polarization vectors satisfy the conditions,
$$ {\varepsilon}_\pm ^{\mu}({\bf 0}){\varepsilon}_{\pm \mu}({\bf 0}) = 0. $$
This has also been observed  in \cite{tyutin}, although its origin remained slightly mysterious. Here however, it is possible to interpret these conditions as a
consequence of the usual orthogonality relations,
$$ {\varepsilon}_+^{\ast \mu}({\bf 0}){\varepsilon}_{- \mu}({\bf 0}) = 0. $$
together with the parity law (2.2.17). 
These observations suggest an inbuilt doublet structure in the MCSP model.
The embedded doublet structure, related by the augmented parity transformations, in the 
MCSP theory will be further explored in the next subsection.
\subsection{An alternative approach: $U(1)$ invariance and doublet structure}
The above methods of calculating the polarization vectors depend on the
 existence of a rest frame. The results obtained in this frame are Lorentz
boosted to an arbitrary moving frame. There is another approach which directly
yields the polarization vectors from a solution of the free field equations
of motion. We now discuss this and compare with the previous analysis.

Let us consider the MCS theory (with $\vartheta > 0$). Since it has a
single mode
of mass $\vartheta$, it is possible to write a general expression for 
the polarization vector, satisfying the Lorentz gauge condition (2.1.6)
and the equation of motion (2.1.7),
\bigskip
$$ \eta_{\mu} = N \left(k_{\mu} - g_{\mu 0}\frac{{\vartheta}^2}{\omega} - i \frac{\vartheta}{\omega} \epsilon_{\mu \alpha 0} k^{\alpha} \right) \eqno {(2.3.1)} $$
with, $\omega = k_0 = \sqrt{{\vartheta}^2 + |{\bf k}|^2}$ and $N$ is the 
normalization. This is fixed from the condition ($\eta^{\ast \mu} \eta_{\mu} = -1)$,
\bigskip
$$ N = \frac{1}{\sqrt{2}} \frac{\omega}{\vartheta |{\bf k}|} $$
This expression for the polarization vector was given in \cite{tyutin}. However it 
looks quite different from the previous result (2.1.22b). To make an effective
comparison, let us pass to the rest frame.  Here we face a difficulty. 
In component form, (2.3.1) may
be expressed as,
$$ \eta_{\mu} = \frac{1}{\sqrt{2}\vartheta} \left(|{\bf k}|, \frac{\omega}{|{\bf k}|}
(k_1 + i \frac{\vartheta}{\omega}k_2), \frac{\omega}{|{\bf k}|}(k_2 -i\frac{\vartheta}{\omega}k_1)\right) \eqno {(2.3.2)} $$
It is seen that $\eta_0 = 0 $ in the rest frame $(k_{\mu} = (\vartheta, 0, 0))$, where $\vartheta > 0 $ only is being considered.
 But $\eta_1$ and $\eta_2$ do not have a smooth limit. 
To achieve this we exploit the $U(1)$ invariance of $\eta_{\mu}$ by introducing a phase angle $\phi$,
\bigskip 
$$ k_1 = |{\bf k}| \cos \phi $$
$$ k_2 = |{\bf k}| \sin \phi \eqno{(2.3.3)} $$
Then, since $\vartheta = \omega$ in the rest frame, we find, 
\bigskip
$$ \eta_{\mu} = \frac{1}{\sqrt{2}}(0, e^{i\phi}, -ie^{i\phi}) \eqno{(2.3.4)} $$
Up to a $U(1)$ phase, this exactly coincides with (2.1.14), thereby proving 
the equivalence of the two results. 

Following identical techniques the polarization vectors for MCSP theory turn
out as,
\bigskip
$$ \varepsilon_{\pm \mu} = 
\frac{1}{\sqrt{2}}\frac{\omega_{\pm}}{|{\bf k}|\sqrt{\omega_{\pm}^2 - |{\bf k}|^2}} \left(k_{\pm \mu} - g_{\mu 0}\frac{\omega_{\pm}^2 - |{\bf k}|^2}{\omega_{\pm}} - i \frac{\omega_{\pm}^2 - |{\bf k}|^2 - m^2}{\theta \omega_\pm}
\epsilon_{\mu \alpha 0} k^{\alpha}\right) \eqno{(2.3.5)} $$
where, $\omega_\pm = \sqrt{\theta_{\pm}^2 + |{\bf k}|^2}$.
Once again this does not have a smooth transition to the rest frame. But 
we can show its equivalence with the expressions (2.2.15) by adopting the previous
technique. Expressions similar to (2.3.5) were reported earlier in\cite{tyutin}.

Different representations of the polarization vectors find uses in different contexts. For instance, the expressions (2.3.1) and (2.3.5) are useful\cite{tyutin} in
 analyzing the fermion mass variance in MCS theory. On the other hand, the
 expressions given in (2.1.22b) and (2.2.16) reveal the presence of a doublet
structure in the MCSP model. 
Specifically, the pair of polarization vectors $ {\varepsilon}_{\pm}^{\mu}$, 
corresponding to the distinct massive modes ${\theta}_{\pm}$, can be exactly 
identified with the polarization vectors for a doublet of MCS models,
$$ {\cal L}_+ = -{\frac{1}{4}}{F^{\mu \nu}(A)F_{\mu \nu}(A)} + {\frac{{\theta
}_+}{2}}{\epsilon}^{\mu \nu \lambda}A_{\mu}{\partial}_{\nu}A_{\lambda} \eqno{(2.3.6a)}  $$

$$ {\cal L}_- = -{\frac{1}{4}}{F^{\mu \nu}(B)F_{\mu \nu}(B)} - {\frac{{\theta
}_-}{2}}{\epsilon}^{\mu \nu \lambda}B_{\mu}{\partial}_{\nu}B_{\lambda}. \eqno{(2.3.6b)}  $$

The necessary symmetry features are preserved provided both ${\theta}_{\pm} 
> 0 $ or ${\theta}_{\pm} < 0 $. It then follows from  (2.1.22b) that the
polarization vectors of the MCS doublet exactly match with (2.2.15). The two
massive modes $\theta_{\pm}$ of the doublet are exactly identified with
the pair found in the MCSP model.

Yet another way of understanding the doublet structure is to look at the 
$m^2 \rightarrow 0$ limit of the MCSP model (2.2.1), which then reduces to 
the MCS model. From (2.2.11a,b) we see that this limit corresponds to two
possibilities;
\bigskip
$$ (i) \hskip 1.0 cm \theta_+ \rightarrow 0; \theta \rightarrow -\theta_- \eqno{(2.3.7a)} $$
$$ (ii) \hskip 1.0 cm \theta_- \rightarrow 0; \theta \rightarrow \theta_+ \eqno{(2.3.7b)} $$
These two cases $(\theta \rightarrow \pm \theta_\pm)$ exactly correspond 
to the MCS doublet (2.3.6a, b). Likewise the polarization vectors(2.2.15) also
map to the corresponding doublet structure. Note that this expression is 
divergent for $\theta_+ \rightarrow 0$ or $\theta_- \rightarrow 0$, but this
mode does not correspond to the physical scattering amplitude when fermions
are coupled \cite{tyutin}.

It is worthwhile to mention that the limit $m^2 \rightarrow 0$ takes a second
class system  to a first class one. From the view point of a constrained system,
 such a limit is generally not smooth. However, the polarization vector shows
a perfectly smooth transition. We might also recall that the $m^2 \rightarrow 0$ limit in the second class Proca model, to pass to the Maxwell theory, is 
problematic due to the change in the nature of the constraints. This is also
manifested in the structure of the polarization vectors. Setting the $m^2 \rightarrow 0$ limit in  
the relevant expressions for the Proca model does not yield the Maxwell
theory polarization vector. In this way, therefore, the massless limit in the MCSP theory is quite distinctive. Since a pair of MCS theories get mapped to the
MCSP theory, such a smooth transition exists.  This also shows the reason that the 
infrared singularities in the MCS model can be cured by including a mass term(MCSP model)
and taking the vanishing mass limit\cite{tyutin}.

Another point worth mentioning is that, if $\theta_+ = \theta_-$, then 
$\theta = 0$  from (2.2.11a). This means that a doublet of MCS theories having
 the same topological mass but with opposite helicities, maps to a massive 
Maxwell model, with mass $m^2 = \theta_+^2 = \theta_-^2$. In this case parity is a symmetry
which is also seen from the generalized transformations(2.2.16). This mapping was
 discussed earlier\cite{deser} in terms of the basic fields of the respective
models.
\section{Polarization vectors in the  Hamiltonian formalism}
The analysis of polarization vectors is now done in the Hamiltonian formulation,
which is based on Dirac's constrained algorithm. Different gauge conditions will be considered. The equivalence with the Lagrangian formulation is clearly seen.
\subsection{Maxwell-Chern-Simons theory}
In this subsection we again consider the MCS field in the Lorentz gauge, but
with the difference that the Lorentz gauge condition is now  imposed at 
the level
of the Lagrangian of the model itself. Consider the Lagrangian,
\bigskip
$$ {\cal L}_{\alpha} = -{\frac{1}{4}}{F^{\mu \nu}F_{\mu \nu}} + {\frac{\vartheta}{2}}
{\epsilon}^{\mu \nu \lambda}A_{\mu}{\partial}_{\nu}A_{\lambda} - \frac{1}{2 \alpha}
(\partial \cdot A)^2   \eqno {(3.1.1a)} $$
which is obtained from (2.1.1) by adding an extra gauge fixing term 
$- \frac{1}{2 \alpha}(\partial \cdot A)^2$. (In this subsection $\alpha$ represents the gauge parameter). For simplicity, the parameter $\vartheta$ is taken 
positive.
If the vector field $A^{\mu}$ satisfies the Lorentz constraint (2.1.3), the 
Lagrangian ${\cal L}_{\alpha}$  is equivalent to the MCS Lagrangian given by (2.1.1). 
The value of the gauge parameter $\alpha$ being
arbitrary, we make the choice $\alpha = 1$(Feynman gauge). With this choice, after an integration 
by parts in the action, ${\cal L}_1$ transforms to,
$$ {\cal L}_1 = -\frac{1}{2} {\partial}_{\mu} A_{\nu} \partial^{\mu} A^{\nu} + \frac{1}{2}\partial_{\mu}\left[ A_{\nu}(\partial^{\nu} A^{\mu}) - (\partial_{\nu} A^{\mu})A^{\mu}\right] + {\frac{\vartheta}{2}} {\epsilon}^{\mu \nu \lambda}A_{\mu}{\partial}_{\nu}A_{\lambda}. $$ 
Ignoring the total divergence term we write,
\bigskip
 $${\cal L}_1 = -\frac{1}{2} {\partial}_{\mu} A_{\nu} \partial^{\mu} A^{\nu} + {\frac{\vartheta}{2}} {\epsilon}^{\mu \nu \lambda}A_{\mu}{\partial}_{\nu}A_{\lambda}.
 \eqno {(3.1.1b)} $$  
 The conjugate momenta are given by, 
\bigskip
$$ \pi^{\mu} = \frac{{\partial}{\cal L}_1}{{\partial}\dot{A}_{\mu}} = (-\dot{A}^0 , -\dot{A}^i + \vartheta \epsilon^{i j} A_j) \eqno {(3.1.2)} $$
The Hamiltonian corresponding to (3.1.1$b$) is 
\bigskip
$$ H_1 = \int d^2 x \left[-{\frac{1}{2}} \pi^{\mu}\pi_{\mu} + {\frac{1}{2}}\partial^kA^\nu \partial_k A_\nu \right]  $$
$$ + \int d^2 x\left[
 - \frac{\vartheta}{2} \epsilon^{i j}\left( A_i \pi_j + A_0 \partial_i A_j + A_i \partial_j A_0 \right) + {\frac{1}{8}}\vartheta^2 {\bf A}^2 \right] \eqno {(3.1.3)} $$
The  Hamilton's equations following from
$$\dot{A}^{\mu} = \{ A^{\mu}, H_1 \}$$
 and 
$$\dot{\pi}^{\mu} =\{ {\pi}^{\mu} , H_1 \}$$
  are explicitly given as follows.
\bigskip
$$ \dot{A}^0 = -\pi^0 \eqno{(3.1.4a)} $$
$$ \dot{A}^1 = -\pi^1 - \frac{\vartheta}{2} A^2 \eqno{(3.1.4b)} $$
$$ \dot{A}^2 = -\pi^2 + \frac{\vartheta}{2} A^1 \eqno{(3.1.4c)} $$
$$ \dot{\pi}^0 = -\nabla^2 A^0 + \vartheta \epsilon^{i j}\partial_i A_j \eqno{(3.1.4d)} $$
$$ \dot{\pi}^1 = -\nabla^2 A^1 - \vartheta \partial^2 A^0 + \frac{\vartheta^2}{4}A^1 -\frac{\vartheta}{2}\pi^2 \eqno {(3.1.4e)} $$ 
$$ \dot{\pi}^2 = -\nabla^2 A^2 + \vartheta \partial^1 A^0 + \frac{\vartheta^2}{4}A^2 + \frac{\vartheta}{2}\pi^1 \eqno {(3.1.4f)} $$
Since our aim is to obtain the explicit form of the polarization vectors of the
field, we consider solutions of the form,
\bigskip
$$ A^{\mu} = \sum_{\lambda = 1}^2\eta^{\mu}(\lambda, {\bf k})a_{{\bf k} \lambda} \exp[ik \cdot x] + c.c \eqno{(3.1.5a)} $$
$$ \pi^{\mu} = \sum_{\lambda = 1}^2 \eta^{\mu}(\lambda, {\bf k})b_{{\bf k} \lambda} \exp[ik \cdot x] + c.c \eqno{(3.1.5b)} $$
Note that the polarization vectors $\eta^{\mu}(k)$ used in the previous section have been expanded in terms of their basis vectors. Since $\eta^{\mu}(k)$ is
space-like there are at most two such linearly independent vectors characterized
by $\lambda$.
The above solutions when substituted in (3.1.4a) and (3.1.4d) give, in the 
rest frame $(k_0, 0, 0)$ of the quanta of excitations,
\bigskip
$$ (ik_0 a_1 + b_1)\eta^0 (1,{\bf 0}) + (ik_0 a_2 + b_2)\eta^0 (2,{\bf 0}) = 0 $$ 
and
$$ (ik_0 b_1)\eta^0 (1,{\bf 0}) + (ik_0 b_2)\eta^0 (2,{\bf 0}) = 0 $$
In these equations, the symbol ${\bf k}$ in $a_{{\bf k} \lambda}$ and $b_{{\bf k} \lambda} $ has been suppressed.
It can be seen that the determinant $(a_1b_2 - a_2b_1)$ of the coefficients 
does not vanish, since in that case $A^{\mu}$ would be proportional to $\pi^{\mu}$. Hence the only solutions to the above set of two equations are the trivial ones.
That is 
\bigskip
$$ \eta^0(\lambda, {\bf 0}) = 0 \eqno {(3.1.6)} $$
Similar substitution of (3.1.5a, b) in (3.1.4b, c, e and f) gives, in the rest 
frame,
\bigskip
$$ {\Sigma}_{MCS(H)} \bar{\eta}(0) = 0 \eqno{(3.1.7)} $$
where
\bigskip
$$ {\Sigma}_{MCS(H)} = \left( \begin{array}{cccc}
  ik_0 a_1 + b_1 & \frac{\vartheta}{2}a_1 & ik_0 a_2 + b_2  & \frac{\vartheta}{2}a_2 \\
  -\frac{\vartheta}{2}a_1 & ik_0 a_1 + b_1 & -\frac{\vartheta}{2}a_2 & ik_0 a_2 + b_2  \\
  ik_0 b_1 - \frac{\vartheta^2}{4}a_1 & \frac{\vartheta}{2}b_1 & ik_0 b_2 - \frac{\vartheta^2}{4}a_2 & \frac{\vartheta}{2}b_2 \\
 -\frac{\vartheta}{2}b_1 & ik_0 b_1 - \frac{\vartheta^2}{4}a_1 & -\frac{\vartheta}{2}b_2 & ik_0 b_2 - \frac{\vartheta^2}{4}a_2 \end{array} \right)  \eqno{(3.1.8)} $$ 
and 
\bigskip
$$ \bar{\eta}(0) = \left( \begin{array}{c}
{\eta}^1(1,{\bf 0}) \\
{\eta}^2(1,{\bf 0}) \\
{\eta}^1(2,{\bf 0}) \\
{\eta}^2(2,{\bf 0})
\end{array} \right)   \eqno{(3.1.9)}   $$
The solutions of (3.1.7) are given by, 
\bigskip
$$   {\eta}^2(\lambda, {\bf 0}) = -i{\eta}^1(\lambda, {\bf 0}) \eqno{(3.1.10)} $$
For $\vartheta > 0,$ (3.1.6) and (3.1.10) agree with the result (2.1.11a,b) obtained in the Lagrangian approach. The agreement clearly will be preserved
for the polarizations vectors in a moving frame also.

Now we show that the above result is a special case of a more general one in which we introduce a 
Nakanishi-Lautrup auxiliary field $\cal B$ in the MCS Lagrangian and linearize the gauge fixing 
term.
In this covariant gauge formalism the Lagrangian (3.1.1$a$) is expressed as,
\bigskip
$$ {\cal {L}}_{\alpha} = -{\frac{1}{4}}{F^{\mu \nu}F_{\mu \nu}} + {\frac{\vartheta}{2}}{\epsilon}^{\mu \nu \lambda}A_{\mu}{\partial}_{\nu}A_{\lambda} + 
{\cal {B}} \partial_{\mu}A^{\mu} +
 {\frac{{\alpha}}{2}} {{\cal {B}}^2} \eqno {(3.1.11)} $$
Notice that ${\cal L}_{\alpha}$ does not 
contain $\dot{\cal B}$ and that it is linear in $\dot{A}^0$. The Euler-Lagrange equations of motion which follow from (3.1.11) are, 
\bigskip
$$ \Box A^{\mu} - \partial^{\mu}(\partial_{\nu} A^{\nu} + {\cal{B}}) + 
\vartheta \epsilon^{\mu \nu \lambda} \partial_{\nu} A_{\lambda} = 0 \eqno {(3.1.12a)} $$
and
$$ {\cal {B}} = - \frac{1}{\alpha} \partial_{\nu}A^{\nu}. \eqno {(3.1.12b)} $$
With the choice ${\alpha} = 1 $ and eliminating $\cal B $ using the
equation of motion (3.1.12b), one can get ${\cal L}_1$. The momenta conjugate to the fields $A^0, A^i$ and $\cal B$ are, respectively, given by
\bigskip
$$ \pi_0 = {\cal{B}}  $$
$$ \pi_i = \partial_i A_0 - \dot{A}_i + \frac{\vartheta}{2} \epsilon_{i j} A^j 
$$
$$ \pi_{{\cal{B}}} = 0 $$
The Hamiltonian obtained from ${\cal {L}}_{\alpha}$ is 
\bigskip
$$ H_{\alpha} = \frac{1}{2}\int d^2 x \left[(\pi_i)^2 + \vartheta \epsilon^{i j}\pi^i A^j + \frac{\vartheta^2}{4} + \frac{1}{2}(F^{i j})^2\right] $$
$$ + \int d^2 x \left[A^0 (\partial^i \pi^i - \frac{\vartheta}{2} \epsilon^{i j}
\partial^i A^j) - {\cal{B}}(\partial^i A^i) - \frac{1}{\alpha} {\cal {B}}^2\right]  
\eqno {(3.1.13)} $$
The constraints of the model are,
\bigskip
$$ \Phi_1 = \pi_0 - {\cal{B}} \approx 0 \eqno {(3.1.14a)} $$
and
$$ \Phi_2 = \pi_{\cal{B}} \approx 0 \eqno {(3.1.14b)} $$
which form a second class pair. Setting the the first constraint strongly equal
to zero, one obtains $\pi_0 = {\cal{B}}$, using which one can eliminate the 
auxiliary field ${\cal{B}}$ from the Hamiltonian:
\bigskip
$$ H_{\alpha} = \frac{1}{2}\int d^2 x \left[(\pi_i)^2 + \vartheta \epsilon^{i j}\pi^i
A^j + \frac{\vartheta^2}{4} + \frac{1}{2}(F^{i j})^2\right] $$
$$ + \int d^2 x \left[A^0 (\partial^i \pi^i - \frac{\vartheta}{2} \epsilon^{i j}
\partial^i A^j) - \pi^0 (\partial^i A^i) - \frac{1}{\alpha} (\pi^0)^2\right]
\eqno {(3.1.15)} $$
The brackets between the fields $A^{\mu}$ and their conjugate momenta 
$\pi_{\nu}$ are given by,
\bigskip
$$ \{ A^{\mu}({\bf x}, t), \pi_{\nu}({\bf y}, t) \} = {g^{\mu}}_{\nu} \delta
({\bf x} - {\bf y}). \eqno {(3.1.16)} $$
It should be mentioned that the Dirac brackets in the ($A^{\mu}, \pi_{\nu}$)
sector are identical to their Poisson brackets.
Hence the Hamilton's equations for $A^{\mu}$ and $\pi^{\mu}$ are obtained from
$$ \dot{A}^{\mu} = \{A^{\mu}, H_{\alpha} \}, \hskip 0.5cm 
 \dot{\pi}^{\mu} = \{ \pi^{\mu},  H_{\alpha} \} $$
and are the following.
\bigskip
$$ \dot{A}^0 = \partial^i A^i - {\alpha} \pi^0 \eqno {(3.1.17a)} $$
$$ \dot{\pi}^0 = \frac{\vartheta}{2} \epsilon^{i j} \pi^i A^j -
 \partial^i \pi^i \eqno {(3.1.17b)} $$
$$ \dot{A}^i = - \pi^i - \frac{\vartheta}{2} \epsilon^{i j} A^j + \partial^i A^0
 \eqno {(3.1.17c)} $$
$$ \dot{\pi}^i = \frac{\vartheta^2}{4}A^i + \partial^j F^{i j} - \frac{\vartheta}{2}\epsilon^{i j} \left(\partial^i A^0 + \pi^j\right) - \partial^i \pi^0 \eqno {(3.1.17d)} $$
Substitution of the expansions (3.1.5a, b) in (3.1.17a, b), in the rest frame
$(k_0, 0, 0)$ leads to, 
\bigskip
$$ \left[ \eta^0(1, {\bf 0})a_1 + \eta^0(2, {\bf 0})a_2 \right]ik_0 + \left[ \eta^0(1, {\bf 0})b_1 + 
\eta^0(2, {\bf 0})b_2 \right]\alpha = 0 $$
and
$$ \left[ \eta^0(1, {\bf 0})b_1 + \eta^0(2, {\bf 0})b_2 \right]ik_0  = 0 $$ 
If $\alpha \neq \infty$ we can rewrite the above set as 
$$\left( \begin{array}{cc}
a_1 & a_2 \\
b_1 & b_2 
\end{array} \right)\left( \begin{array}{c}
 \eta^0(1, {\bf 0}) \\
 \eta^0(2, {\bf 0})
\end{array} \right) = 0 $$
The above equation does not have any nontrivial solution as $a_1b_2 - a_2b_1 \neq 0$ for
the same reason mentioned earlier in the case of $\alpha = 1$. Hence,   
$$ \eta^0(\lambda, {\bf 0}) = 0 $$
if $\alpha$ is finite. When $\alpha = \infty$ there is no unambiguous solution.
This case corresponds to the fact that when $\alpha = \infty$, the gauge
fixing term in $\cal{L}_\alpha$(3.1.1a) vanishes. The expansions (3.1.5a, b)
when substituted in (3.1.17c, d) yields, 
$$ \eta^2(\lambda, {\bf 0}) = -i\eta^1(\lambda, {\bf 0}) $$ 
These results are independent of the gauge parameter and naturally agree with
the previous $\alpha = 1$ calculation.
\subsection{Maxwell - Chern - Simons - Proca theory}
Taking the MCSP Lagrangian (2.2.1) the canonical momenta are defined as,
\bigskip
$$ \pi^i = \frac{{\partial}{\cal L}}{{\partial}\dot{A}_i} = -\left(F^{0 i} + \frac{\theta}{2}\epsilon^{i j}A^{j}\right) \eqno {(3.2.1)} $$
and 
\bigskip
$$ \pi^0 \approx 0 \eqno {(3.2.2)} $$
is the primary constraint.
The canonical Hamiltonian is, 
\bigskip
$$ H_{MCSP} = {\frac{1}{2}}\int d^2 x \left[{{\pi}_i}^2 + \frac{1}{2}{F_{i j}}^2 + \left(\frac{{\theta}^2}{4} + m^2\right){A_i}^2 -
\theta {\epsilon_{i j}}A_i \pi_j + m^2 {A_0}^2\right] $$
$$ + \int d^2 x{A_0}\Omega \eqno {(3.2.3)}   $$
where, 
\bigskip 
$$ \Omega = \partial_i\pi_i - \frac{\theta}{2}\epsilon_{i j}\partial_i A_j -m^2A_0 \approx 0 \eqno {(3.2.4)}   $$
is the secondary constraint. Using (3.2.4) to eliminate $A_0$ from (3.2.3), we obtain the reduced Hamiltonian,
\bigskip
$$ H_R = {\frac{1}{2}}\int d^2 x \left[{{\pi}_i}^2 + \left(\frac{1}{2} + \frac{{\theta}^2}{8m^2}\right){F_{i j}}^2 +
 (\frac{{\theta}^2}{4} + m^2){A_i}^2 - \theta\epsilon_{i j}A_i\pi_j\right]  $$  
$$  + {\frac{1}{2m^2}}\int d^2 x \left[(\partial_i \pi_i)^2 - \theta \partial_i \pi_i \epsilon_{l m} \partial_l A_m \right]. \eqno {(3.2.5)}  $$
The only  non vanishing bracket between the phase space variables is,
\bigskip
$$ \{A^i({\bf x},t), \pi^j({\bf y},t)\} = - \delta^{i j}\delta({\bf x} - {\bf y})  \eqno {(3.2.6)}  $$
Therefore the Hamilton's equations are given by 
\bigskip
$$ \dot{A}^i = \{ A^i, H_R \}  
             = -\pi^i + \frac{1}{m^2}\partial^i\partial^j\pi^j -  \frac{\theta}{2}\epsilon^{i j}A^j 
- \frac{\theta}{2m^2}\epsilon^{l m}\partial^i\partial^lA^m \eqno {(3.2.7)} $$
 and $$ \dot{\pi}^i = \{ \pi^i, H_R \}
 = \left(1 + \frac{\theta^2}{4m^2}\right)\left[\partial^i\partial^jA^j - \partial^j\partial^jA^i + m^2A^i\right]
-\frac{\theta}{2}\epsilon^{i j}\left[\pi^j + \frac{1}{m^2}\partial^j\partial^k\pi^k\right] \eqno {(3.2.8)} $$
We consider solutions (in terms of the polarization vectors $\varepsilon({\bf k}, \lambda))$ of the form,
\bigskip 
$$ A^i = \sum_{\lambda = 1}^2 \varepsilon^i(\lambda, {\bf k})a_{{\bf k} \lambda} \exp[ik \cdot x] + c.c \eqno{(3.2.9)} $$  
$$ \pi^i = \sum_{\lambda = 1}^2 \varepsilon^i(\lambda, {\bf k})b_{{\bf k} \lambda} \exp[ik \cdot x] + c.c \eqno{(3.2.10)} $$
Substitution of the above solutions in the Hamilton's equations (3.2.7) and (3.2.8)
yields, respectively,
\bigskip
$$ \sum_{\lambda = 1}^2 -ik_0\varepsilon^i(\lambda, {\bf k})a_{{\bf k} \lambda} = \sum_{\lambda = 1}^2  \{[\varepsilon^i(\lambda, {\bf k}) +  
{\frac{1}{m^2}}k^ik^j\varepsilon^j(\lambda, {\bf k})]b_{{\bf k} \lambda}   $$
$$ + {\frac{\theta}{2}}
[\epsilon^{i j}\varepsilon^j(\lambda, {\bf k}) - {\frac{1}{m^2}}\epsilon_{l m}k^ik^l\varepsilon^m(\lambda, {\bf k})]
a_{{\bf k} \lambda} \} \eqno{(3.2.11)} $$
and 
\bigskip
$$ \sum_{\lambda = 1}^2 -ik_0\varepsilon^i(\lambda, {\bf k})b_{{\bf k} \lambda} = \sum_{\lambda = 1}^2 \{[k^ik^j
\varepsilon^{j}(\lambda, {\bf k}) - k^jk^j\varepsilon^{i}(\lambda, {\bf k}) - m^2 \varepsilon^{i}(\lambda, {\bf k})]
(1 + \frac{\theta^2}{4m^2})a_{{\bf k} \lambda} $$
$$ + \frac{\theta}{2}\epsilon^{i j}[\varepsilon^{j}(\lambda, {\bf k}) - {\frac{1}{m^2}}k^jk^k\varepsilon^{k}(\lambda,
 {\bf k})]b_{{\bf k} \lambda} \}  \eqno{(3.2.12)}   $$
In the rest frame $(k^0, 0, 0)$, the above set of four equations can be written in the matrix form as, 
\bigskip
$$ {\Sigma}_{MCSP(H)}\bar{\varepsilon} = 0  \eqno{(3.2.13)}   $$
where
$$ {\Sigma}_{MCSP(H)} = $$ 
$$ \left( \begin{array}{cccc} 
  b_1 + ik_0a_1 & \frac{\theta}{2}a_1 & b_2 + ik_0a_2 & \frac{\theta}{2}a_2  \\
  -\frac{\theta}{2}a_1 & b_1 + ik_0a_1 & -\frac{\theta}{2}a_2 & b_2 + ik_0a_2 \\
  -Dm^2a_1 + ik_0b_1 & \frac{\theta}{2}b_1 &  -Dm^2a_2 + ik_0b_2 & \frac{\theta}{2}b_2 \\
  -\frac{\theta}{2}b_1 & -Dm^2a_1 + ik_0b_1 & -\frac{\theta}{2}b_2 & -Dm^2a_2 + ik_0b_2 
\end{array} \right)  \eqno{(3.2.14)}   $$
and 
\bigskip
$$ \bar{\varepsilon} = \left( \begin{array}{c}
{\varepsilon}^1(1,{\bf 0}) \\
{\varepsilon}^2(1,{\bf 0}) \\
{\varepsilon}^1(2,{\bf 0}) \\
{\varepsilon}^2(2,{\bf 0}) 
\end{array} \right)   \eqno{(3.2.15)}   $$
with $D= (1 + \frac{\theta^2}{4m^2}) $.
For a nontrivial solution of (3.2.13), $ \det{\Sigma}_{MCSP(H)} = 0. $ This condition, after a straightforward 
algebra, reduces to 
\bigskip
$$ (a_1b_2 - a_2b_1)^2\left[k_0^4 -2k_0^2(\frac{\theta^2}{2} + m^2) + m^4\right] = 0 \eqno {(3.2.16)} $$
from which it follows that 
\bigskip
$$ \left[ k_0^4 -2k_0^2(\frac{\theta^2}{2} + m^2) + m^4 \right] = 0 \eqno {(3.2.17)} $$
That is, 
\bigskip
$$ k_0 = \sqrt{\frac{\theta^2}{4} + m^2} \pm \frac{\theta}{2} = \theta_\pm \eqno {(3.2.18)} $$
Replacing $k_0$ with $\theta_+$ in (3.2.13) we obtain, after suitable
manipulations, the following relationship between the components of
$\vec{\varepsilon}(\lambda, {\bf 0})$;
\bigskip
$$ \varepsilon^2(\lambda, {\bf 0}) = -i\varepsilon^1(\lambda, {\bf 0}) $$
For $k_0 = \theta_- $ the corresponding relationship is given by
$$ \varepsilon^2(\lambda, {\bf 0}) = +i\varepsilon^1(\lambda, {\bf 0}). $$
Denoting the the rest frame polarization vectors corresponding
to $ k_0 = \theta_\pm $ by $\varepsilon_\pm({\bf 0})$, the above two
expressions can be written as,
$$ \varepsilon^2_\pm({\bf 0}) = \mp i \varepsilon^1_\pm({\bf 0}). \eqno {(3.2.19)} $$
which agrees with the relationship obtained from the  Lagrangian
formalism. The polarization vectors in moving frames are obtained
by  boosting the rest frame vectors appropriately, and the result
obviously agrees with that obtained earlier within the Lagrangian framework.

\section{Wigner's little group in 2+1 dimensions}
MCS theory provides an example in 2+1 dimensions, where massive excitations
can co-exist with gauge invariance, as we have mentioned earlier.
Here the mass is induced by the topological CS term 
right at the classical(tree) 
level, where the dual $\tilde{F}_\mu $ of the field strength tensor,
$(\tilde{F}_\mu = \frac{1}{2}\epsilon_{\mu \nu \lambda}F^{\nu \lambda})$
can be shown to satisfy a Klein-Gordon type equation with the mass given
by the Chern - Simons parameter\cite{djt}. In this sense the model is different from
the Schwinger model, where the mass for the gauge field is generated at 
the one loop quantum level.

From a constrained Hamiltonian analysis of the model\cite{djt}, one can
of course show that it admits a  first class constraint which acts as
a generator of $ U(1)$ gauge transformation just like the pure Maxwell theory.
On the other hand, as has been shown by Weinberg\cite{weinberg} and others\cite{han}, 
the translation like generators of  Wigner's little group E(2) for the
massless particles in 3+1 dimensions also act as generators of gauge 
transformation in the pure Maxwell theory. As Maxwell theory only admits   
massless quanta(photons), one cannot go to the rest frame. But one can go 
to a frame where a photon is propagating, say, in the z-direction. 
Wigner's little group E(2), which is a subgroup of the Lorentz group 
SO(1,3), preserves this four momentum $k^\mu$; but the polarization vector
$\xi^\mu (k)$ undergoes the gauge transformation, 
$$ \xi^{\mu}(k) \rightarrow \xi^{\prime \mu}(k) = \xi^\mu(k) + f(k)k^\mu $$
under the the action of `translation-like' generators of E(2). One can therefore ask whether the same argument is valid in 2+1 dimensional
Maxwell theory as well, where the little group is isomorphic to ${\cal {R}} {\times}{\cal {Z}}_{2}$\cite{binegar}. 
Furthermore, whether can this little group generate gauge 
transformation in MS theory also, where the quanta are massive and
a rest frame is available? We try to analyze these issues in this section.
To put the thing in a clear perspective, we shall also analyze the free
Maxwell theory and the MCS theory side by side for ready comparison between these two 
theories. 

To begin with let us provide a brief derivation of the exact form of
Wigner's little group for massless particles in 2+1 dimensions. For that we essentially follow
\cite{qft}.

Let  $k^\mu = \left( \begin{array}{c}
                    1 \\
                    0 \\
                    1
\end{array} \right)$ be a light - like vector with the spatial component
directed to the y-axis.
Let ${W^{\mu}} _{\nu}$ be an arbitrary element of the little group.
Then we must have, 
\bigskip
$$ (Wk)^\mu = {W^{\mu}}_{\nu} k^\nu = k^\mu   \eqno {(4.1)} $$
Also if $t^\mu = \left( \begin{array}{c}
                    1 \\
                    0 \\
                    0
\end{array} \right)$ is a unit time-like vector then we must have 
$t^{\mu}t_{\mu} = 1$ and $t^{\mu}k_{\mu} = 1$. It follows immediately, using 
(4.1) that, 
\bigskip
$$ (Wt)^{\mu} (Wt)_{\mu} = t^{\mu} t_{\mu} = 1 \eqno {(4.2a)} $$
$$ (Wt)^{\mu} k_{\mu} = (Wt)^{\mu} (Wk)_{\mu} =  t^{\mu} k_{\mu} = 1 \eqno {(4.2b)} $$
Any 3-vector$(Wt)^{\mu}$ satisfying (4.2b) can be written as, 
$$ (Wt)^\mu = \left( \begin{array}{c}
              1 + \zeta \\
                  \alpha \\
                  \zeta
  \end{array} \right) \eqno {(4.3)} $$
Upon using the normalization condition (4.2a) one gets $\zeta = \frac{\alpha^2}{2}
$
so that (4.3) reduces to, 
\bigskip
$$ (Wt)^\mu = \left( \begin{array}{c}
               1 + \frac{\alpha^2}{2} \\
               \alpha \\ 
               \frac{\alpha^2}{2}
\end{array} \right). \eqno {(4.4)} $$
Thus we see that the "Minkowski" scalar product between the time-like 
3-vector $t^\mu$ and light-like vector $k^\mu$ and their Lorentz invariance 
(4.2b) suggest the general form (4.3) for $ (Wt)^\mu$ with two unknown real
parameters $\alpha$ and $\zeta$.
And then the normalization condition of $ t^\mu $ and its Lorentz invariant
form (4.2a) determine one of the unknowns $\zeta $
in terms of $\alpha $ to yield (4.4). One can proceed similarly for other two
unit space-like vectors $ s^\mu = \left( \begin{array}{c}
                                  0 \\
                                  0 \\
                                  1 
\end{array} \right) $ and $ r^\mu = \left( \begin{array}{c}
                                  0 \\
                                  1 \\
                                  0
\end{array} \right) $, to find, 
\bigskip
$$ (Ws)^\mu = \left( \begin{array}{c}
              \frac{\alpha^\prime}{2} \\
              \alpha^\prime \\
              1 - \frac{\alpha^\prime}{2}
\end{array} \right) \eqno {(4.5)} $$ 
and
\bigskip
$$ (Wr)^\mu = \left( \begin{array}{c}
              c \\
              a \\
              c
\end{array} \right);  a = \pm 1  \eqno {(4.6)} $$
These unknown parameters $\alpha^\prime$ and $c$ can be fixed in terms of $\alpha$
using the orthonormality relations and their Lorentz invariances, 
\bigskip
$$ (Wt)^{\mu}(Ws)_{\mu} = t^{\mu}s_{\mu} = 0 $$ 
$$ (Ws)^{\mu}(Wr)_{\mu} = s^{\mu}r_{\mu} = 0 \eqno {(4.7)} $$
to get, 
$$ \alpha^\prime = - \alpha  $$
$$ c = a\alpha \eqno {(4.8)} $$ 
We therefore have, 
\bigskip
$$ (Ws)^\mu = \left( \begin{array}{c}
              -\frac{\alpha^2}{2} \\
              -\alpha \\
              1- \frac{\alpha^2}{2}
\end{array} \right) \eqno {(4.9a)} $$
and 
\bigskip
$$ (Wr)^\mu = \left( \begin{array}{c}
             a\alpha \\
              a \\
              a\alpha 
\end{array} \right) \eqno {(4.9b)} $$
Clearly the column matrices $(Wt)^{\mu}, (Wr)^{\mu}$ and $(Ws)^{\mu}$correspond to the three columns of the  ${W^{\mu}}_{\nu}(\alpha)$ matrix and making use of
(4.4) and (4.9), it can be written as, 
\bigskip
$$  {W^{\mu}}_{\nu} (\alpha) = \left( \begin{array}{ccc}
  1 + \frac{\alpha^2}{2} & a\alpha & -\frac{\alpha^2}{2} \\
  \alpha & a & -\alpha \\
  \frac{\alpha^2}{2} & a\alpha & 1- \frac{\alpha^2}{2}
\end{array} \right) \eqno {(4.10)} $$
At this stage one can easily show that, 
\bigskip
$$ W(\alpha)\cdot W(\beta) = W(\alpha + \beta) \eqno {(4.11)} $$ 
Thus this one parameter Wigner group is isomorphic to ${\cal R} \times {\cal Z}_{2}$
($\cal R $ is the additive group of real numbers). The $ {\cal Z}_{2}$ factor is     
required to take into account that the value of $a$ is restricted to $\pm1$ (4.6).

The generator G in this representation is clearly given(with $a = +1$)  by,
\bigskip
$$ G = \frac{\partial W}{\partial \alpha}\mid_{\alpha = 0} = 
\left( \begin{array}{ccc}
       0 & 1 & 0 \\
       1 & 0 & -1 \\
       0 & 1 & 0 
\end{array} \right) \eqno {(4.12)} $$
satisfying,
\bigskip
$$ G^2 = \left( \begin{array}{ccc}
          1 & 0 & -1 \\
          0 & 0 & 0 \\
          1 & 0 & -1 
\end{array} \right);   G^3 = 0 \eqno {(4.13)} $$
so that $W(\alpha)$ can be re-expressed as, 
\bigskip
$$ W(\alpha) = e^{\alpha G} = 1 + \alpha G + \frac{1}{2}\alpha^2 G^2 \eqno {(4.14)} $$
Another representation of this group is given by, 
\bigskip
$$ \bar{W}(\alpha) = \left( \begin{array}{cc}
          1 & \alpha \\
          0 & 1 
\end{array} \right) \eqno {(4.15)} $$
Acting on a column matrix $ \left( \begin{array}{c}
           x \\
           1
\end{array} \right) $ (where $x$ is the transverse direction to the momentum) 
it generates translation $x \rightarrow x + \alpha $.
The group of such translations is clearly isomorphic to ${\cal R}$.
The corresponding generator is, 
                               $$\bar{G} = \left( \begin{array}{cc}
                                         0 & 1 \\
                                         0 & 0 
                                     \end{array} \right)$$.

\subsection{Role of the little group as a generator of  gauge transformation}
Let us briefly review the role of $W(\alpha)$ as a generator of gauge
transformation in Maxwell theory\cite{weinberg,han} before taking up the MCS case. For that 
consider a photon of energy $\omega$ to be moving in the $y$-direction and 
polarized in the $x$-direction, so that the potential 3-vector takes the
form 
\bigskip
$$ A^{\mu}(x) = \xi^{\mu}_x \exp(-i(\omega k).x) = \xi^{\mu}_xe^{-i\omega (t - y)} \eqno {(4.1.1a)}  $$
where 
\bigskip
$$ \xi^{\mu}_x = \left( \begin{array}{c}
                             0 \\
                             1 \\
                             0
\end{array} \right) \eqno {(4.1.1b)} $$
is the polarization vector and the subscript $x$ denotes the fact that this
 vector is in the $x$-direction. Also $(\omega k)^{\mu} = p^\mu $ is
the energy-momentum 3-vector in this case. Under the action of $W$ (4.10),
$ \xi^{\mu}_x $ undergoes the transformation,
\bigskip
$$ \xi^{\mu}_x \rightarrow \xi^{\prime \mu}_x = {W^{\mu}}_{\nu}
\xi^{\nu}_x = \xi^{\mu}_x + \alpha k^\mu \eqno {(4.1.2a)} $$
This can be identified as the gauge transformation as the corresponding gauge field undergoes the transformation,
\bigskip
$$ A^\mu (x) \rightarrow  A^{\prime \mu} (x) =A^\mu (x) +  \partial^{\mu}(\frac{i \alpha}{\omega} e^{-i \omega k.x}) \eqno {(4.1.2b)} $$
In contrast the MCS quanta are massive and the polarization vector for
$\vartheta < 0$ takes a simple form, 
\bigskip
$$ \eta^\mu ({\bf 0}) = \frac{1}{\sqrt{2}}\left( \begin{array}{c}
                                0 \\
                               1 \\
                               i
\end{array} \right) \eqno {(4.1.3)} $$
in the rest frame. But note that it has complex entries having both $x$ and
$y$ components unlike the Maxwell photon polarization $\xi^{\mu}_x $.
In fact in their coulomb gauge analysis Devecchi et. al.\cite{girotti} have pointed out 
that the spin $(\frac{\vartheta}{|\vartheta|})$ of the MCS quanta stems from this
particular complex structure  of the polarization vector. We shall also try
to provide a heuristic derivation of spin for MCS quanta by exploiting this
particular complex structure of the polarization vector in the next subsection. 

We shall now investigate whether this same little group can generate similar
gauge transformation on the MCS polarization vector (4.1.3). The question is 
interesting in its own right, as we have explained earlier, because this
little group $W$ corresponds to a massless particle and generates gauge transformation acting on the polarization vector of a massless particle like that of
a photon. Will it generate similar gauge transformation for massive
excitations? To that end, let us apply $W(\alpha)$ on $\eta^\mu ({\bf 0})$
(4.1.3). We find that it undergoes the following transformation, 
\bigskip
$$ \eta^\mu ({\bf 0}) \rightarrow \eta^{\prime \mu} ({\bf 0}) = {W^{\mu}}_{\nu}(\alpha)\eta^\nu ({\bf 0}) =   \frac{1}{\sqrt{2}}\left( \begin{array}{c}
   \alpha - \frac{i}{2}\alpha^2 \\
    1 - i\alpha \\
   \alpha + i(1- \frac{\alpha^2}{2})
\end{array} \right) \eqno {(4.1.4)} $$
Clearly this can not be cast in the form of (4.1.2a). One cannot therefore 
interpret this transformation as a gauge transformation. However, taking  
advantage of the fact that this little group involves a single parameter
only, we can easily construct a (non-unique) representation which does the
required job. This is given by, 
\bigskip
$$ {\tilde{W}^{\mu}}_{\nu}(\alpha) = \left( \begin{array}{ccc}
       1 & \alpha & -i\alpha \\
       0 & 1 & 0 \\
       0 & 0 & 1 
\end{array} \right); \alpha \in {\cal R} \eqno {(4.1.5)} $$
so that in place of (4.1.4) one has
\bigskip
$$ \eta^\mu ({\bf 0}) \rightarrow \eta^{\prime \prime \mu}({\bf 0}) \equiv
 {\tilde{W}^{\mu}}_{\nu}  \eta^\nu ({\bf 0}) = 2\alpha k^\mu + \eta^\mu ({\bf 0}) \eqno {(4.1.6)} $$
where $k^\mu = \left( \begin{array}{c}
 1 \\
 0 \\
 0
\end{array} \right)$ and modulo the mass factor this is the energy-momentum
     3-vector of a massive particle in the rest frame. This is clearly the
 desired form like (4.1.2a). One can easily show that $\tilde{W}(\alpha) \cdot \tilde{W}(\beta) = \tilde{W}(\alpha + \beta) $ and there exists a natural isomorphism
between $W(\alpha)$ and $\tilde{W}(\alpha)$, so that $\tilde{W}(\alpha)$ also
furnishes a representation of the little group. We have thus shown that 
the little group for a massless particle, in an appropriate representation,
can generate a gauge transformation on the polarization vector of the massive
MCS quanta in its rest frame (4.1.6). The little group for the massive particle,
which in this case can be  trivially seen to be $O(2)$, however, does not have any
role as a generator of gauge transformation. This does not mean however that
they are completely unrelated. In fact, 
one can write $W(\alpha)$ for $|\alpha| < 1$, as a product of three matrices;
\bigskip
$$ W(\alpha) = B^{-1}_y(\alpha) R(\alpha) B_x(\alpha)  \eqno {(4.1.7)} $$
where 
\bigskip
$$ B^{-1}_y(\alpha) =  \left( \begin{array}{ccc}
 \frac{2 - \alpha^2}{2\gamma} & 0 & -\frac{\alpha^2}{2\gamma} \\
 0 & 1 & 0 \\
 -\frac{\alpha^2}{2\gamma} & 0 & \frac{2 - \alpha^2}{2\gamma}
\end{array} \right);  R(\alpha) = \left( \begin{array}{ccc}
 1 & 0 & 0 \\
 0 & \gamma & -\alpha \\
 0 & \alpha & \gamma 
\end{array} \right); 
 B_x(\alpha) = \left( \begin{array}{ccc}
\frac{1}{\gamma} & \frac{\alpha}{\gamma} & 0 \\
 \frac{\alpha}{\gamma} & \frac{1}{\gamma} & 0 \\
0 & 0 & 1 
\end{array} \right) \eqno {(4.1.8)} $$
with $ \gamma = \sqrt{1 - \alpha^2}$. These matrices are themselves the elements
of the Lorentz group $SO(1,2)$; $ B_x$ represents a boost along the $x$-direction,
$R$ represents a  spatial rotation in the $x-y$ plane and $B^{-1}_y$ 
represents a boost along the negative $y$-direction. Appropriate 
transformations in this order can preserve the energy-momentum 3-vector 
of a particle moving in the $y$-direction. Here $R$ clearly corresponds to the
little group of a massive particle. 
Thus (4.1.7)  relates the elements of the connected parts of identity element
of  the little group of massless particles with massive ones as long as 
$|\alpha| < 1 $. But this does not provide the natural homomorphism existing
between ${\cal R} $(the additive group of real numbers) with $SO(2)$.

Coming back to the issue of similarities and dissimilarities between the 
polarization vectors of pure Maxwell theory and that of MCS theory, note
that a photon state is entirely characterized by (4.1.1b), where both the 
`spatial' transversality condition ${\bf k} \cdot \vec{\xi}_x = 0$
and the `temporal' gauge condition $\xi^{0}_x = 0$ are satisfied.
Thus the gauge field configuration (4.1.1a) corresponds to the radiation gauge.
Clearly the same gauge condition will no longer be valid under a Lorentz boost.
However, as we show following \cite{han,bc} that the radiation gauge condition can 
still be satisfied, provided the gauge field undergoes an appropriate 
gauge transformation preceding the Lorentz boost.

Using (4.1.2b), the gauge transformed field configuration $A^{\prime \mu}(x)$
can be written as, 
\bigskip
$$ A^{\prime \mu}(x) = A^{\mu}(x) + \alpha k^\mu e^{-i\omega(t - y)} = A^{\mu}(x) + \alpha k^\mu e^{-i p \cdot x} \eqno {(4.1.9)} $$ 
for a photon of energy $\omega$ and propagating in the $y$-direction. A Lorentz
boost of velocity $\tanh \phi $ for example, in the $x$-direction yields,
\bigskip
$$ \tilde{A}^{\prime \mu} = \left( \begin{array}{ccc}
 \cosh \phi & \sinh \phi & 0 \\
 \sinh \phi & \cosh \phi & 0 \\
 0 & 0 & 1 
\end{array} \right) \left( \begin{array}{c} \alpha \\ 1 \\ \alpha 
\end{array} \right) e^{-ip^{\prime} \cdot x^{\prime}} 
 = \left( \begin{array}{c} \alpha \cosh \phi + \sinh \phi \\ \alpha \sinh \phi + \cosh \phi \\
\alpha \end{array} \right) e^{-ip^{\prime} \cdot x^{\prime}} \eqno {(4.1.10)} $$
where $p^{\prime \mu} $ is the appropriate energy-momentum 3-vector in the
new co-ordinate frame $x^{\prime \mu}$ and is given by, 
\bigskip 
$$ p^{\prime \mu} = \omega \left( \begin{array}{c}
  \cosh \phi \\
  \sinh \phi \\
    1
\end{array} \right)   \eqno {(4.1.11)}  $$
Preservation of spatial transversality condition implies that we must have, 
\bigskip 
$$ {\tilde{\bf A}}^{\prime}(x^{\prime}) \cdot {\bf p}^{\prime} = \omega
\left( \begin{array}{c}
\cosh \phi + \alpha \sinh \phi \\
\alpha 
\end{array} \right)^{T} 
\left( \begin{array}{c}
\sinh \phi \\
1 
\end{array} \right) = 0 \eqno {(4.1.12)} $$
Solving for $\alpha$, one gets, 
\bigskip
$$ \alpha = -\tanh \phi \eqno {(4.1.3a)} $$
This solution, when substituted back in (4.1.10) yields,
\bigskip
$$ \tilde{A}^{\prime 0} (x^{\prime}) = 0 \eqno {(4.1.13b)} $$
which is nothing but the temporal gauge condition. Thus with an appropriate
gauge transformation preceding a Lorentz boost the radiation gauge condition
can be satisfied. But, as we shall see now,  the same is not true for 
MCS Theory. 
Upon gauge transformation,
the polarization vector(4.1.3), in the rest frame, becomes 
\bigskip
$$ \eta^\mu (x) \rightarrow \tilde{\eta}^\mu (x) = \tilde{W}_0
(\alpha) \eta^\mu (x) = 
 \frac{1}{\sqrt{2}} \left( \begin{array}{ccc}
 1 & \alpha & -i\alpha \\
 0 & 1 & 0 \\
 0 & 0 & 1 
\end{array} \right) \left( \begin{array}{c} 
0 \\
1 \\
i
\end{array} \right)      
 = \frac{1}{\sqrt{2}}\left( \begin{array}{c}
2 \alpha \\
1 \\
i 
\end {array} \right)  \eqno {(4.1.14)} $$
Then a Lorentz boost along $x$-axis, for example transforms this to,
\bigskip
$$ \tilde{\eta}^\mu \rightarrow {\tilde{\eta}}^{\prime \mu} = \frac{1}{\sqrt{2}}
\left( \begin{array}{c}
\sinh \phi + 2 \alpha \cosh \phi \\
\cosh \phi + 2 \alpha \sinh \phi \\
i 
\end{array} \right) \eqno {(4.1.15)} $$
Simultaneously, the $k^\mu =\left( \begin{array}{c}
1 \\
0 \\
0
\end{array} \right)$, associated to the rest frame, transforms  to, 
\bigskip
$$ k^\mu \rightarrow k^{\prime \mu} = \left( \begin{array}{c}
\cosh \phi \\
\sinh \phi \\
0 
\end{array} \right) \eqno {(4.1.16)} $$
Now the spatial transversality condition ${\bf k} \cdot {\bf A} = 0 $
is trivially satisfied in the rest frame. Demanding that the same condition is 
satisfied in the boosted frame as well, one gets using (4.1.15) and (4.1.16),
\bigskip
$$ \left( \begin{array}{c}
\sinh \phi \\
0 
\end{array} \right)^T \left( \begin{array}{c}
\cosh \phi + 2 \alpha \sinh \phi \\
i
\end{array} \right) = 0,  \eqno {(4.1.17)} $$
which when solved for the gauge transformation parameter $\alpha$  in terms of the 
boost parameter $\phi$, yields, 
\bigskip
$$ \alpha = - \frac{1}{2 \tanh \phi}. \eqno {(4.1.18)} $$
So just like in the Maxwell case the spatial transversality condition can be 
maintained in any boosted frame, provided the boost is preceded by a suitable gauge transformation. However the time component $\tilde{\eta}^{\prime 0}$ does not vanish for,
$\alpha$ satisfying (4.1.18):
\bigskip
$$ {\tilde{\eta}}^{\prime}_ 0 = -\frac{1}{\sqrt{2}\sinh \phi} \eqno {(4.1.19)} $$  
This is in contrast with the Maxwell case (4.1.13b). Nevertheless, $ \tilde{\eta}^{\prime 0}$
(4.1.19) can be made to vanish in the limit $\phi \rightarrow \infty$ i.e. when $\alpha 
\rightarrow -\frac{1}{2} $ and $\tanh \phi \rightarrow 1$.
 This is precisely the case when velocity approaches the speed of light. 

For arbitrary $\phi$, one can write,
\bigskip
$$ \tanh \phi = \frac{|k^1|}{k^0} \eqno{(4.1.20)} $$

Using (4.1.18),  (4.1.20) and the mass-shell condition $p^2 = \vartheta^2$, one can simplify (4.1.15) to get the polarization vector as,  
\bigskip
$$ \tilde{\eta}^\mu = \frac{1}{\sqrt{2}}\left( \begin{array}{c}
-\frac{\vartheta}{|k^1|} \\
0 \\
i
\end{array} \right) \eqno {(4.1.21)}  $$
This is nothing but the polarization vector of MCS theory in the Coulomb gauge,
as has been calculated by Devecchi et. al.\cite{girotti} for a boost along $x$-axis.
\subsection{Little group and spin of MCS quanta}
We conclude this section by making certain observations and try to
provide a heuristic derivation of the spin of MCS quanta. Note that the norm of the MCS polarization vector is defined as $\eta^{\ast \mu} \eta_{\mu}$ which
determines whether this polarization vector  is space-like or time-like. The presence of the complex conjugation indicates that the norm is formally invariant under the group $U(2,1)$
which contains the Lorentz group $O(2,1)$ as a subgroup. The form of the corresponding
little group, the counterpart of (4.10), can be easily obtained by following the same method to get,  
\bigskip
$$ {\cal W}(\phi, \alpha, \zeta) =  \left( \begin{array}{ccc}
1+\zeta & e^{-i\phi} \alpha^{\ast} & -\zeta \\
\alpha & e^{-i\phi} & -\alpha \\
\zeta & e^{-i\phi} \alpha^{\ast}  & 1 - \zeta
\end{array} \right) \eqno {(4.2.1)} $$
subject to the constraint, 
\bigskip
$$ \zeta + \zeta^\ast = |\alpha|^2 \eqno{(4.2.2)} $$
Note that this form is associated with the massless particles moving in the $y$-direction and thus preserves $k^{\mu} = \left( \begin{array}{c} 
1 \\
0 \\
1
\end{array} \right)$ i.e., $(Wk)^{\mu} = k^{\mu}$, just as in (4.1). 
This can be factorized as 
\bigskip
$$ {\cal W}(\phi, \alpha, \zeta) = {\cal W}(0, \alpha, \zeta)  \left( \begin{array}{ccc}
1 & 0 & 0 \\
0 & e^{-i\phi} & 0 \\
0 & 0 & 1
\end{array} \right) \eqno {(4.2.3)} $$

This is reminiscent of how the corresponding element of the 
 Wigner's little group $\subset O(3,1)$ in 3+1 dimensions can be  factorzed  leaving an $SO(2)$ factor on the right, 
the eigenvalue of which determines the helicity\cite{qft}. The question that 
naturally arises is 
whether this second factor in (4.2.3) is anyway related to the spins of the MCS quanta?
We shall now try to provide certain plausible arguments to answer this question.
First note that the second factor in (4.2.3) is an element of $U(1)$ group,
which is again isomorphic to the group $SO(2)$, the little group for {\it
massive} particles  like MCS quanta.

Now considering the case $\vartheta < 0$ of the doublet(2.1.12),  the space part $\vec{\eta}({\bf 0})$ of the polarization vector $\eta^{\mu}({\bf 0})
(= \left( \begin{array}{c}
  0 \\
  \vec{\eta}({\bf 0})
\end{array} \right)$ can be written as,  
\bigskip
$$ \vec{\eta}({\bf 0}) =  \left( \begin{array}{c}
  {\eta}^1({\bf 0}) \\
  {\eta}^2({\bf 0})
\end{array} \right) =\frac{1}{\sqrt{2}}  \left( \begin{array}{c}
1 \\
i \end{array} \right) = {\eta}^1({\bf 0}) e_1 + {\eta}^2({\bf 0}) e_2 \eqno {(4.2.4)} $$
with $e_1 = \left( \begin{array}{c}
1 \\ 
0
\end{array} \right)$ and $e_2 = \left( \begin{array}{c}
0 \\
1
\end{array} \right)$ now correspond to the plane polarized basis. As is done for Maxwell theory in $3+1$ dimensions\cite{qft}, here too we can go to the circular- polarized basis, 
\bigskip
$$ e_+ = \frac{1}{\sqrt{2}}(e_1 + ie_2) = \frac{1}{\sqrt{2}}\left( \begin{array}{c} 
1 \\
i
\end{array} \right); $$
$$e_- = e_+ ^\ast = \frac{1}{\sqrt{2}}(e_1 - ie_2) = 
\frac{1}{\sqrt{2}} \left( \begin{array}{c}
1 \\
-i
\end{array} \right) \eqno{(4.2.5)} $$
One can easily check the orthonormality of the basis($<e_\pm, e_\pm> = e^{\dagger}_{\pm} e_{\pm} = 1; <e_+,
e_-> = e^{\dagger}_+ e_- = 0$ ). $\vec{\eta}({\bf 0})$ can now be expressed in this basis as, 
\bigskip
$$ \vec{\eta}({\bf 0}) = \eta^+ ({\bf 0})e_+ + \eta^- ({\bf 0})e_- \eqno {(4.2.6a)} $$
where the corresponding coefficients $\eta^{\pm}({\bf 0})$ can be easily seen to be given by, 
\bigskip
$$ \eta^+ ({\bf 0}) = \frac{1}{\sqrt{2}}(\eta^1 ({\bf 0}) - i \eta^2({\bf 0})) = 1  $$
$$ \eta^- ({\bf 0}) =  \frac{1}{\sqrt{2}}(\eta^1 ({\bf 0}) + i \eta^2({\bf 0})) = 0 \eqno {(4.2.6b)} $$
Thus in this basis $\vec{\eta}({\bf 0})$ is just $e_+$, with $e_-$ 
disappearing completely. It will be just the opposite for $\vartheta > 0$ case. 
On the other hand, the elements of the little group $SO(2)$ for massive 
particles, $ R(\phi) = \left( \begin{array}{cc}
\cos \phi & \sin \phi \\
-\sin \phi & \cos \phi
\end{array} \right)$ can be represented in a diagonal form $R_d(\phi)$ in the $(e_+, e_-)$ basis as,
\bigskip
$$ R_d(\phi) =  \left( \begin{array}{ccc}
1 & 0 & 0 \\
0 & e^{-i\phi} & 0 \\
0 & 0 & e^{i\phi} 
\end{array} \right) = \left( \begin{array}{ccc}
1 & 0 & 0 \\
0 & e^{-i\phi} & 0 \\
0 & 0 & 1
\end{array} \right) \left( \begin{array}{ccc}
1 & 0 & 0 \\
0 & 1 & 0 \\
0 & 0 & e^{i\phi}
\end{array} \right) \eqno {(4.2.7)} $$ 
where the first factor $\left( \begin{array}{ccc}
1 & 0 & 0 \\
0 & e^{-i\phi} & 0 \\
0 & 0 & 1
\end{array} \right)$ is the second factor of (4.2.3) and is associated with
the massless particle moving in the $y$-direction. The second factor 
$\left( \begin{array}{ccc} 
1 & 0 & 0 \\
0 & 1 & 0 \\
0 & 0 & e^{i\phi}
\end{array} \right) $ similarly can be made to be associated to the massless 
particle moving in the $x$-direction. The massive MCS quanta in its rest frame
 can not have anything to do with either first or second factor in (4.2.7) 
individually. From symmetry considerations, the product of both of these 
commuting factors should be taken in to account.The correct representation should therefore be their product $R_d(\phi)$(4.2.7). This also has the desired 
property that det$R(\phi) = 1$, so that we are back to the group $SO(2)$
just as in 3+1 dimensional Maxwell theory. 
Clearly, the basis vector $e_+$ picks up a phase $e^{-i\phi}$ under 
rotation as can be seen by applying $R^T(\phi)$ on $e_1 $ and $e_2$.
 Since the rotation 
operator is given by $R(\phi) = e^{i\phi J}$ ($J$ being the angular momentum
 operator), it follows that $J$ has eigenvalue $-1$ in this case. Proceeding similarly 
for the other doublet($\vartheta > 0$), the eigenvalue will be $+1$. 

One can thus note that although Wigner's little group for massless particles(4.10)
does not have an $SO(2)$ subgroup unlike its (3+1) dimensional counterpart\cite{weinberg},
one can consider a bigger group $U(2,1)$ containing the Lorentz group as a subgroup and isolate a factor
isomorphic to $SO(2)$. Doing it for massless particles moving in 
$x$ and $y$ directions respectively, one can take their product to
construct the desired representation $R_d(\phi)$(4.2.7), which is 
indeed the diagonal representation of $SO(2)$ -the little group for  massive 
MCS quanta in ($e_+, e_-$) basis. As happens for MCS theory $\eta^2({\bf 0})$ and $\eta^1({\bf 0})$ are not
independent unlike Maxwell case in 3+1 dimensions. Their particular relation
in turn implies that $\eta^-({\bf 0}) \left( \eta^+({\bf 0}) \right)$ vanishes for $\vartheta < 0  
\left(\vartheta > 0\right)$ case allowing only one spin state for   MCS quanta
with spin (-1) or (+1) respectively.

\section{Conclusion}
A detailed analysis of the polarization vectors in planar field theories
involving both a topological mass and a usual mass has been done. The
structure of these vectors is crucial for the reduction formulae, the
study of unitarity in topologically massive gauge theories\cite{tyutin}, as
well as the massless limit of these theories augmented by a normal mass term.
While some analysis of the polarization vectors has been done\cite{tyutin,
girotti}, it does not reveal the various subtleties and nuances.

Our general approach using either Lagrangian or Hamiltonian techniques,
has shown a $U(1)$ invariance (in the k-space) in the form of the 
polarization vectors. This is quite distinct from the usual Abelian
invariance associated with gauge theories. The $U(1)$ invariance 
reported here is connected with the presence of the Chern-Simons(CS)
term and has nothing to do with the presence or absence of gauge freedom
in the action. The CS term leads to complex entries in the polarization 
vector, thereby manifesting a $U(1)$ invariance. This can be contrasted with
the pure Maxwell theory where all entries are real so that there is no
$U(1)$ invariance of this type. The complex structure of the polarization
vectors provided an alternative understanding of the co-existence of
gauge invariance with mass. It was seen that the massive modes were
physical while the massless ones could be gauged away and hence were
unphysical. This is the exact counter part of the Maxwell theory where
the roles of the massive and massless modes are reversed.

 As a further
application, the $U(1)$ invariance was exploited to show the equivalence
between the results quoted in the literature\cite{tyutin,girotti}  
and those found by us. Additionally, the latter naturally revealed
a mapping between the Maxwell-Chern-Simons-Proca(MCSP)  model and a doublet
of Maxwell-Chern-Simons theories with opposite helicities. This was also
helpful in studying the massless limit of the MCSP model.

The difference in the manifestation of gauge invariance in the usual 
Maxwell theory and the Maxwell-Chern-Simons theory prompted us to
study Wigner's little group in 2+1 dimensions. Indeed, as is known,
this group acts as a generator of gauge transformation 
in the Maxwell theory in 3+1 dimensions. We found the explicit structure 
of the  little group in 2+1 dimensions. It was shown how different 
representations of this group act as a generator of gauge transformations
in either the MCS theory or the Maxwell theory. 
This is quite distinctive because in one case the modes are massive, while in the other these are massless.
Finally based on the structure of the little group, a heuristic derivation 
of the spin of the MCS quanta was presented.

\end{document}